\documentclass[aps,prl,twocolumn,showpacs,superscriptaddress,groupedaddress]{revtex4}  
\usepackage{graphicx}  
\usepackage{dcolumn}   
\usepackage{bm}        
\usepackage{amssymb}   
\usepackage{amsmath}

\begin{document}


\title{Description of  shape coexistence in $^{96}$Zr based on the collective quadrupole Bohr Hamiltonian}

\author{D.A. Sazonov}%
\author{E.A. Kolganova}
\address{Joint Institute for Nuclear Research, 141980 Dubna, Moscow region, Russia}
\address{Dubna State University, 141982 Dubna, Moscow Region, Russia}
\author{T.M. Shneidman}
\address{Joint Institute for Nuclear Research, 141980 Dubna, Moscow region, Russia}
\address{Kazan Federal University, Kazan 420008, Russia}
\author{R.V. Jolos}
\address{Joint Institute for Nuclear Research, 141980 Dubna, Moscow region, Russia}
\address{Dubna State University, 141982 Dubna, Moscow Region, Russia}
\author{N. Pietralla}
\author {W.Witt}
\address{Institut f\"ur Kernphysik, TU Darmstadt, Schlossgartenstr. 9, D-64289, Darmstadt, Germany}
\date{\today}

\begin{abstract}
\begin{description}
\item[Background:] Experimental data on $^{96}$Zr indicate coexisting spherical and deformed structures with small mixing amplitudes.
\item[Purpose:] To investigate the properties of the low-lying collective states of $^{96}$Zr  based on the
collective quadrupole  Bohr Hamiltonian.
\item[Method:] The $\beta$-dependent collective potential having two minima -- spherical and deformed,  is fixed
so to describe experimental data in the best way.
\item[Results:]  Good agreement with the experimental data on the excitation energies, $B(E2)$ and $B(M1)$ reduced transition probabilities is obtained.
\item[Conclusion:] It is shown that the low-energy structure of $^{96}$Zr can be reproduced in a satisfactory way in the geometrical model with a potential function supporting shape coexistence. However, the excitation energy of the $2^+_2$ state can be reproduced only if the rotation inertia coefficient is taken five times smaller then the vibrational one in the region of the deformed well. It is shown also that shell effects are important for the description of the $B(M1;2^+_2 \rightarrow 2^+_1)$ value. An indication on the influence of the pairing vibrational mode on the $\rho^2 (0^+_2 \rightarrow 0^+_1)$ value is obtained.
\end{description}
\end{abstract}

\maketitle

\section{Introduction}
Shape coexistence in nuclei is a remarkable phenomenon which became a widespreaded feature that may occur in many nuclei.  The occurrence of different shapes has its origin in the evolution of shell structure with excitation energy and varying occupation of nucleonic orbitals~\cite{Wood92,Casten,Heyde11}. A theoretical approach suitable for consideration of the phenomena related to nuclear shape dynamics is a collective nuclear model.
The basic idea of the nuclear collective model is that even though such a quantum many body system as the atomic nucleus is characterized by a huge amount of microscopic degrees of freedom they are organized in collective modes playing a crucial role in determining nuclear structure. As a next step, a collective Hamiltonian can be constructed which includes as the basic ingredients the deformation-dependent collective potential and the tensor of inertia ~\cite{Li}. This Hamiltonian determines nuclear collective dynamics.

A shape evolution of the nuclear states can happen as a function of excitation energy, angular momentum and the number of nucleons. The shape transition with the number of nucleons has been noticed in many chains of isotopes and isotones but appears to be rather gradual in many cases. The abrupt change of shape in Zr isotopes~\cite{Cheifetz,Federman78,Federman79} is exceptional. This is a notable feature of nuclear structure in $A \sim 100$ nuclei. The structure of Zr isotopes has been studied within the framework of many different models
~\cite{Sieja,Garcia,Boyukata,Liu,Petrovici11,Petrovici12,Rodrigues,Skalski,Xiang,Mei,Skalski97,Ozen,Fortune}. Strongly deformed states coexisting with a nearly spherical ground state have been reported for Sr and Zr isotopes~\cite{ISOLDE,Buscher,Lherson}. In parallel with a smooth or abrupt establishment of a deformed shape a significant mixing between the configurations of a different shape or a suppressed mixing of two such configurations can be considered. Such information can be obtained from electromagnetic transition probabilities. By measuring the electromagnetic decay properties of the collective $2^+_2$ state of
$^{96}$Zr~\cite{Kremer} the high purity of the coexisting states has recently been established.

It was discussed in ~\cite{Chakreborty} that
the occurrence of the shape coexistence can be related to the existence of a sufficiently large energy gap between subshells. Closely spaced subshells lose their individuality due to pairing correlations and behave as a single, large subshell supporting deformations. In the presence of well-defined subshells in the spherical single particle level scheme  a strong redistribution of nucleons over the single particle levels can take place by particle-hole transitions  with increasing excitation energy. This helps to stabilize the deformation of the excited states when the ground state is spherical~\cite{Togashi}. Substantial change of a configuration of nucleons leads to a large difference in microscopic structure between the spherical states and the states with deformation-optimized shell structure. This property can decrease mixing of two configurations of different shape.

Some low-lying excited states of $^{96}$Zr are related by strong $E2$ transitions. Therefore, it is natural to consider them in the framework of the collective quadrupole model with the  Bohr Hamiltonian. However, the appearance of  shape coexistence is due to  shell effects. Therefore, shell effects are important for the consideration of the structure of $^{96}$Zr and it is interesting to study to what extent the structure of $^{96}$Zr can be described in the framework of the collective geometrical model with phenomenologically determined potential, and how the impact of shell effects are manifested in such a description.

The aim of the present paper is to investigate  an option to explain the properties of the low-lying collective $0^+_{1,2}$ and $2^+_{1,2}$ states of $^{96}$Zr and a weak mixing of the configurations characterized by spherical and deformed shapes based on the Bohr collective quadrupole Hamiltonian. At the end of the paper we also address briefly the characteristics of the
4$^+_1$ and 0$^+_3$ states.

\section{Model}

The general Bohr collective quadrupole Hamiltonian takes the form~\cite{Niksic}
\begin{eqnarray}
\label{eq:one}
	H = -\dfrac{\hbar^2}{2B_0} \dfrac{1}{\sqrt{w r}} \dfrac{1}{\beta^4} \dfrac{\partial}{\partial\beta} \beta^4
	\sqrt{\dfrac{r}{w}} b_{\gamma\gamma} \dfrac{\partial}{\partial\beta}  \cr
	+ \hat{T}_{\beta\gamma} + \hat{T}_\gamma + \dfrac{\hbar}{2B_0} \sum_{\kappa} \dfrac{\hat{I}^2_\kappa}{\Im_\kappa} + V(\beta).
\end{eqnarray}
The first term in~(\ref{eq:one}) presents the kinetic energy of $\beta$-vibrations. The second and the third terms are  connected to the $\gamma$-degree of freedom. The last two terms in~(\ref{eq:one}) are the rotational and potential energies, correspondingly. Above
\begin{eqnarray}
\label{eq:two}
	w &=& b_{\beta\beta} b_{\gamma\gamma} - b^2_{\beta\gamma},  \nonumber \\
	r &=& b_1 b_2 b_3 ,\\
	\Im_\kappa &=& 4 b_\kappa \beta^2 \sin^2 (\gamma - \dfrac{2\kappa\pi}{3}). \nonumber
\end{eqnarray}

The parameter $B_{0}$ is a common dimensional scaling factor for the components of the inertia tensor. Thus, the coefficients $b_{\beta\beta}$, $b_{\gamma\gamma}$, $b_{\beta\gamma}$ and $b_{\kappa}$ are dimensionless inertia coefficients for the $\beta$- and $\gamma$- vibrations, and the rotational motion. The collective potential is determined below.

For keeping out task at a manageable size we assume in this paper that the $\gamma$-degree of freedom can be separated from $\beta$  in the case of
the low-lying states of $^{96}Zr$.  It means that a schematic form of the potential $V(\beta,\gamma)=V(\beta)+V_{\gamma}(\gamma)$ is used. However, it is not enough for complete separation of $\beta$ and $\gamma$ degrees of freedom since they are coupled through the kinetic energy term. As it was shown in \cite{Caprio} this coupling provides a shift of the probability distributions in $\beta$ of the various states toward larger values of $\beta$. This evolution of the probability distribution can be understood qualitatively as a five-dimensional analog of the centrifugal stretching which effectively produces an additional $\sim 1/\beta^2$ type term in $V(\beta)$.
Strictly  speaking,  this term is different for different states. However, this difference can be imitated  by the anharmonic correction to the spherical part of the total collective potential as it influences differently the wave functions of different states.
Thus, the potential obtained below fitting the experimental data includes, in fact, the effect of the $\beta-\gamma$ coupling through the kinetic energy term. Thus, we put below the coefficient $b_{\beta\gamma}=0$  and consider $b_{\gamma\gamma}$ as a constant.
We assume also, for simplicity, that $b_1=b_2=b_3\equiv b_{\rm rot}$. Additionally we assume that $V_{\gamma}$ provides stabilization around $\gamma=0$. We mention, however, that the results for the potential energy surface of $^{96}$Zr presented in \cite{Kremer,Togashi}, rather indicate a  triaxial shape of $^{96}$Zr in its $2^+_2$ and $0^+_2$ states. Triaxiality should decrease significantly the results presented in Table I for the B(E2;$2^+_2\rightarrow 2^+_1$) value and the quadrupole moment $Q(2^+_2)$. Thus, the experimental information on these quantities will give us information on the shape of $^{96}$Zr.

Separating the $\beta$-degree of freedom,  we obtain as a result the following Hamiltonian which will be used to analyze the apparent shape coexistence phenomena in $^{96}$Zr:
\begin{eqnarray}
\label{eq:three}
	H &=& -\dfrac{\hbar^2}{2B_0} \frac{1}{\sqrt{b_{\beta\beta}b^3_{\rm rot}}} \dfrac{1}{\beta^4} \dfrac{\partial}{\partial\beta} \beta^4
	\sqrt{\dfrac{b^3_{\rm rot}}{b_{\beta\beta}}} \dfrac{\partial}{\partial\beta} \\
	&+& \dfrac{\hbar}{2B_0}  \dfrac{\hat{\vec I}^2 - \hat{I_3}^2}{3 b_{\rm rot} \beta^2} + V(\beta). \nonumber
\end{eqnarray}

The volume element in the collective space is
\begin{eqnarray}
\label{eq:four}
 d \tau_{coll} =  d \beta \beta^4 \sqrt{b_{\beta\beta}b^3_{\rm rot}}  d \Omega.
\end{eqnarray}

In cases when the $\beta$-dependence of $b_{\beta\beta}$ and $b_{\rm rot}$ is important  it is convenient to introduce the following notations
\begin{eqnarray}
\label{eq:five}
d \alpha (\beta) \equiv b^{1/2}_{\beta\beta} (\beta) d \beta, \quad
\tau^{1/2} (\beta) \equiv \beta^4 b^{3/2}_{\mbox{\rm rot}},
\end{eqnarray}
where $\alpha (\beta)$ is a modified collective variable.
However, as it is found in our calculations the interpretation of the existing experimental data on
$0^+_{1,2}$ and $2^+_{1,2}$ states does not require an introduction of a  $\beta$-dependent $b_{\beta\beta}$ coefficient. Thus, we put $b_{\beta\beta}$=1 and do not use below the modified collective variable $\alpha$.

To exclude from the Hamiltonian a term with the first derivative over $\beta$ let us present the collective wave function $\Psi$ as
\begin{eqnarray}
\label{eq:six}
\Psi (\beta) = g(\beta) \Phi (\beta)
\end{eqnarray}
and determine $g(\beta)$ so that the Schr\"odinger equation for $\Phi(\beta)$ will not include a first derivative of $\Phi$.
Then
\begin{eqnarray}
g(\beta) = \tau^{-1/4} (\beta) \nonumber
\end{eqnarray}
and the Schr\"odinger equation for $\Phi$ takes the form:
\begin{eqnarray}
\label{eq:seven}
\left\{ -\dfrac{\hbar^2}{2B_0} \frac{d^2}{d \beta^2} + \dfrac{\hbar^2}{2B_0}  \dfrac{\hat{\vec I}{\,^2} - \hat{I}_3^{\,2}}{3 b_{\rm rot} \beta^2} +
V(\beta)  \right. \\
\left. + \dfrac{\hbar^2}{2B_0}  \left[ \frac{1}{4 \tau} \frac{d^2\tau}{d \beta^2} - \frac{3}{16}  \left( \frac{1}{\tau}
\frac{d\tau}{d \beta} \right)^2 \right] \right\}
\Phi = E \Phi. \nonumber
\end{eqnarray}
In these notations the matrix elements of an arbitrary operator $\hat Q$ are calculated as
\begin{eqnarray}
\langle i | \hat Q | j \rangle = \int_0^\infty d \beta \Phi^*_i \hat Q \Phi_j  .
\label{eq:eight}
\end{eqnarray}

Analyzing the experimental data, it was found in~\cite{Jolos07,Jolos08}, that in the case of well deformed  axially symmetric nuclei the inertia coefficient for the rotational motion is several times smaller than the inertia coefficient for the  vibrational motion. In a complete correspondence with this result it is shown below that in order to explain the excitation energy of the $2^+_2$ state it is necessary to take $b_{\rm rot}$ several times less than unity. 

\section{Results of calculations}

\begin{figure}[hbt]
%
\includegraphics[angle=0,width=8.cm]{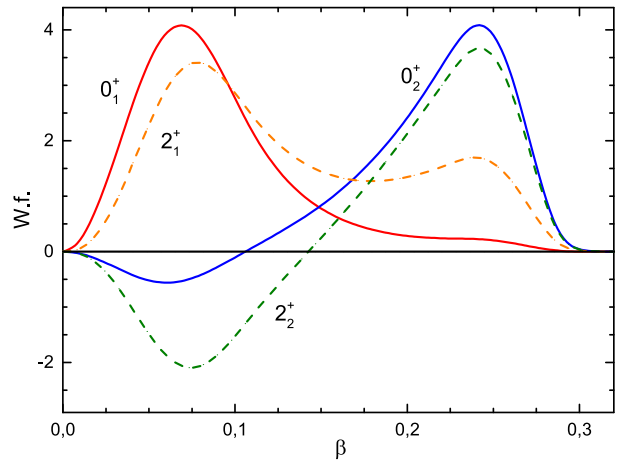}
%
\caption{\label{fig:one}The wave functions of the $0^+_1$, $0^+_2$, $2^+_1$,  and $2^+_2$ states.}
\end{figure}

\begin{figure}[hbt]
%
\begin{center}
\includegraphics[angle=0,width=8.cm]{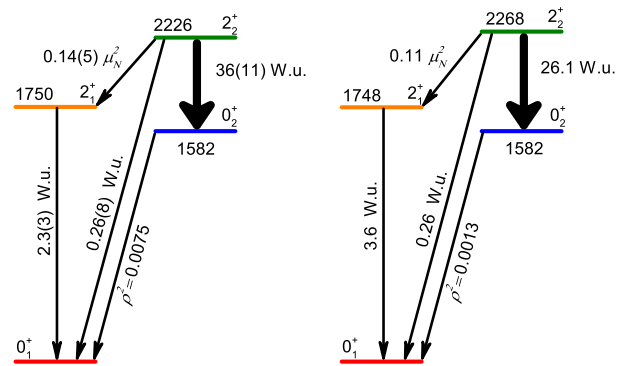}
\end{center}

\caption{\label{fig:two} Experimental (left) and calculated (right) low-lying
$0^+$ and $2^+$ states of $^{96}$Zr. Excitation energies are given in keV.
The values of electric transition probabilities are given in Weiskopf units
and those of magnetic ones are in nuclear magnetons. Experimental data are taken from~\cite{Kremer}.}
\end{figure}

The Hamiltonian~(\ref{eq:seven}) contains two important ingredients which determine the results of our calculations: the potential energy as a function of $\beta$ and the inertia coefficient for the rotational motion. At first, we have tried to determine the shape of the collective potential which allows to get a satisfactory description of the experimental data. Then, the parameter $B_0$ has been varied to fix the energy scale in agreement with the experimental energies. We have initially assumed that the rotational inertia coefficient $b_{\rm rot}$=1.
It was found in this case that the main problem for describing of the experimental data is related to the reproduction of the excitation energy of the second $2^+$ state together with the $0^+_2$ state. This energy spacing was obtained several hundreds keV lower than the experimental value.
As it is shown in Fig.~\ref{fig:one}.
the wave function of the second $2^+$ state is located in the deformed well of the potential.  Therefore, this state can be interpreted qualitatively as a rotational state based on the $0^+_2$ state, and its energy is determined by the deformation at the minimum of the deformed well and the rotational inertia  coefficient. A deformation at the minimum of the deformed well is related to the   $B(E2; 2^+_2\rightarrow 0^+_2$) value. Therefore, the calculated value of the excitation energy of the $2^+_2$ state can be improved only if we assume that the rotational inertia coefficient $b_{\rm rot} < 1 = b_{\beta\beta}$, i.e. smaller than the vibrational inertia coefficient.
This assumption just corresponds to the conclusion obtained in~\cite{Jolos08P,Jolos08,Jolos06,Jolos07} as a result of the analysis of the experimental data on well deformed nuclei. This analysis has indicated that the vibrational inertia coefficient can exceed the rotational one by 4 -- 10 times. The results for the energies of the low-lying collective states and the electromagnetic transition probabilities obtained, assuming that the rotational inertia coefficient is five times smaller than the vibrational one (i.e. $b_{\rm rot}=0.2$) are shown in Table~\ref{table:one} and in Fig.~\ref{fig:two}.

\begin{table}[b]
\caption{The results of calculations of the energies and electromagnetic reduced transition probabilities for $^{96}$Zr. The value of $b_{\rm rot}$ is taken as 0.2. The values of $B(E2)$ are given in Weisskopf units and those of $B(M1)$  in nuclear magnetons.
The value of $Q(2^+_2)$ is given in $e\cdot barn$. The excitation energies are given in keV. The experimental energy of the $0^+_2$ state is used to fix the value of $B_0$. Experimental data are taken from Refs.~\cite{Kremer,Witt}}

\label{table:one}
\begin{center}
\begin{tabular}{l|c|c}
\hline
Energies and transitions						&	calc	& exp \\
\hline
$E(2^+_1)$													&	1748				&	1750		\\
$E(2^+_2)$													&	2268				&	2226		\\
$E(0^+_2)$													&	1582$^*$		&	1582		\\
$B(E2; 2^+_2 \rightarrow 0^+_2)$		&	26.1				&	36(11)	\\
$B(E2; 2^+_1 \rightarrow 0^+_1)$		&	3.6					&	2.3(3)	\\
$B(E2; 2^+_2 \rightarrow 0^+_1)$		&	0.26				&	0.26(8)	\\
$\rho^2(0^+_2 \rightarrow 0^+_1)$		&	0.0013			&	0.0075	\\
$B(E2; 2^+_2 \rightarrow 2^+_1)$		&	2.25				&	$2.8^{+1.5}_{-1.0}	$\\
$B(E2; 2^+_1 \rightarrow 0^+_2)$    & 6.8         & ---     \\
$B(M1; 2^+_2 \rightarrow 2^+_1)$		&	0.11				&	0.14(5)	\\
$Q(2^+_2)$                              &  -0.51                &   ---         \\
\hline
\end{tabular}
\end{center}
\end{table}

The results presented in Table~\ref{table:one} are obtained assuming $\gamma = 0$. This assumption does not influence the results for $B(E2; 2^+_i\rightarrow 0^+_j$). However, in the case of $\gamma = 30^{\circ}$  $Q (2^+_2) = 0$.
As it is seen from the results presented in Table~\ref{table:one} the agreement between the calculated results and the experimental data is satisfactory. The results  for $\rho^2(0^+_2 \rightarrow 0^+_1)$ and $B(M1; 2^+_2 \rightarrow 2^+_1)$ are discussed below.
These quantitites can be calculated using the wave functions shown in Fig.~\ref{fig:one}.  The value of
$\rho^2(0^+_2 \rightarrow 0^+_1)$  is a factor 3.6 smaller than the experimental value.
But both, the experimental and the calculated $\rho^2(0^+_2 \rightarrow 0^+_1)$  values are small in comparison to the corresponding quantitites in other nuclei~\cite{Wood99}.
Using the standard definition of the $E0$ transition operator in the collective model
\begin{eqnarray}
\label{eq:nine}
\hat T (E0) = \frac{3}{4\pi} Ze R^2 \beta^2,
\end{eqnarray}
where $R$ is the nuclear radius, we obtain
\begin{eqnarray}
\label{eq:ten}
\rho^2 (0^+_2 \rightarrow 0^+_1) = \left(\frac{3}{4\pi} Z \right)^2 |\langle 0^+_2 |\beta^2 | 0^+_1 \rangle |^2.
\end{eqnarray}
In order to derive an alternative expression for $\langle 0^+_2 | \beta^2 | 0_1 \rangle$ in terms of other  quantities whose values are known from the experiment let us calculate the double commutator $[[H, \beta^2], \beta^2]$ using the Hamiltonian~(\ref{eq:seven}). The result is
\begin{eqnarray}
\label{eq:eleven}
\left[ \left[ H, \beta^2 \right], \beta^2 \right] = \frac{4\hbar^2}{B_0} \beta^2.
\end{eqnarray}
Taking the average of~(\ref{eq:eleven}) over the ground state $0^+_1$ and assuming that the ground state is mainly related by $E0$ transition to the $0^+_2$ state we obtain
\begin{eqnarray}
\label{eq:twelve}
| \langle 0^+_2 |\beta^2 | 0^+_1 \rangle |^2 \le \frac{2\hbar^2}{B_0} \langle 0^+_1 |\beta^2 | 0^+_1 \rangle \frac{1}{E(0^+_2)},
\end{eqnarray}
where $E(0^+_2)$ is the excitation energy of the second $0^+$ state. The inequality in~(\ref{eq:twelve}) appears because we neglect the contribution to the value of $\langle 0^+_1 | \beta^2 H \beta^2 | 0^+_1 \rangle$ of the $0^+$ states higher in energy than $0^+_2$.
The quantity $\langle 0^+_1 | \beta^2 | 0^+_1 \rangle$ can be expressed with a good accuracy through $B(E2; 2^+_1 \rightarrow 0^+_1)$ using the collective model definition of the $E2$ transition operator $Q_{2\mu}=\dfrac{3}{4}{\pi}eZR^2{\alpha}_{2\mu}$, where assuming axial symmetry we use $\alpha_{2\mu}=D^2_{\mu0}\beta$~\cite{JvonBr}:
\begin{eqnarray}
\label{eq:thirteen}
 \langle 0^+_1 |\beta^2 | 0^+_1 \rangle = \displaystyle{\frac{5 B(E2; 2^+_1 \rightarrow 0^+_1)}{ \left(\frac{3}{4\pi} Ze R^2 \right)^2}   }.
\end{eqnarray}
Substituting~(\ref{eq:twelve}) and~(\ref{eq:thirteen}) into~(\ref{eq:ten}) we obtain
\begin{eqnarray}
\label{eq:fourteen}
\rho^2 (0^+_2 \rightarrow 0^+_1) \le \frac{\hbar^2}{B_0}  \frac{1}{E(0^+_2)} \displaystyle{\frac{10 B(E2; 2^+_1 \rightarrow 0^+_1)}{ e^2 R^4} }.
\end{eqnarray}
In our calculations the value of $\dfrac{\hbar^2}{B_0}$ was previously fixed as  $8.062$ keV in order to reproduce the experimental value of $E(0^+_2)$. Substituting this value  and the experimental values of $E(0^+_2)$ and $B(E2; 2^+_1 \rightarrow 0^+_1)$ into (\ref{eq:fourteen}) we obtain that
\begin{eqnarray}
\label{eq:fifteen}
\rho^2 (0^+_2 \rightarrow 0^+_1) \le 0.0035
\end{eqnarray}
in satisfactory correspondence to the result given in Table~\ref{table:one}.

This result means that the experimental value of $\rho^2(0^+_2 \rightarrow 0^+_1)$ which is $0.0075$  we can reproduce  using the collective model with quadrupole degree of freedom only within a factor of two. We cannot exclude that the pairing vibrational mode plays an important role in the description of the $E0$ transitions~\cite{Iwasaki}.

Consider now the result obtained for the $B(M1; 2^+_2 \rightarrow 2^+_1)$ value. In the collective quadrupole model the part of the $M1$ operator which generates transitions between the states has the form~\cite{Grechukhin}
\begin{eqnarray}
\label{eq:sixteen}
(M1)_\mu = \sqrt{\frac{3}{4\pi}} \frac{e \hbar}{2Mc} \frac{Z}{A} \frac{10}{7\sqrt{\pi}} ( I\alpha_2 )_{1\mu}
\end{eqnarray}
where $I$ is the angular momentum operator. Using the operator~(\ref{eq:sixteen}) we obtain that
\begin{eqnarray}
\label{eq:seventeen}
B(M1; 2^+_2 \rightarrow 2^+_1) \approx 0.03 \cdot  10^{-3} \mu_N^2,
\end{eqnarray}
where $\mu_N$ is a nuclear magneton. This value is more than three orders of magnitude  smaller than the experimental value. The reason for this result is the following. The reduced matrix element of the operator (\ref{eq:sixteen}) is equal to
\begin{eqnarray}
\label{eq:eighteen}
\langle 2^+_2 || M1|| 2^+_1 \rangle = \mu_N \frac{Z}{A} \frac{15}{7 \pi} \int d\beta \Phi_{2^+_2} (\beta) \beta  \Phi_{2^+_1}.
\end{eqnarray}
As it is seen from Fig.~\ref{fig:one} the wave function of the $2^+_2$ state has different signs in the spherical and deformed parts of the potential, whereas the wave function of the $2^+_1$ state does not change sign. As a result the integral in~(\ref{eq:eighteen}) takes a small value. To solve the problem it was suggested in~\cite{Kremer} to use the following $M1$ transition operator:
\begin{eqnarray}
\label{eq:nineteen}
(M1)_\mu =  \mu_N  \sqrt{\frac{3}{4\pi}} g_R (\beta) I_\mu,
\end{eqnarray}
where the $g_R$-factor varies from the shell model value $-0.26$ in the spherical well to $\dfrac{Z}{A}$ in the deformed well with a narrow transition region in $\beta$ where the wave function of the $2^+_2$ state changes its sign.
With this transition operator we obtain $B(M1; 2^+_2 \rightarrow 2^+_1) = 0.11 \mu_N^2$ which  coincides in the limit of the experimental errors with  the experimental value $0.14 \mu_N^2$~\cite{Kremer}. This result stresses the importance of the shell effects for the description of the shape coexistence phenomena, at least, in the case of $^{96}$Zr.

\begin{figure}[hbt]
%
\begin{center}
\includegraphics[angle=0,width=8.cm]{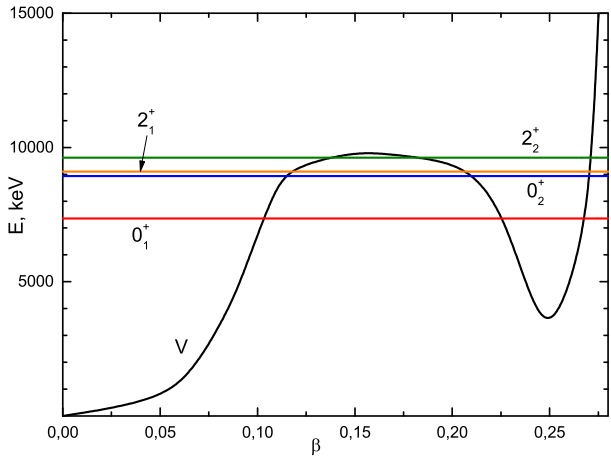}
\end{center}
\caption{\label{fig:three}The phenomenologically assumed potential energy $V(\beta)$ and the calculated energy levels.}
\end{figure}

The collective potential which has been fixed phenomenologically  so to describe the experimental data in the best way is shown in Fig.~\ref{fig:three}.  The  height of the barrier calculated from the position of the ground state is equal to $2.45$ MeV.

The wave functions of the ground and excited states are shown in Fig.~\ref{fig:one}. Their distribution between the spherical and deformed parts of the total potential is characterized by the values obtained by integration of the squares of the wave functions over the regions inside spherical or deformed wells. These values are presented in Table~\ref{table:two}.
\begin{table}[htb]
\caption{\label{table:two}Distribution of the wave function of the $0^+_1$, $0^+_2$, $2^+_1$,  and $2^+_2$ states between the  spherical and deformed parts of the total potential.}
\begin{tabular}{l|c|c|c|c}
\hline
Potential well	&	$0^+_1$ & $0^+_2$ & $2^+_1$ & $2^+_2$ \\
\hline
spherical &	98.9\% & 3.2\% & 77.3\% & 23.7\%		\\
deformed & 1.1 \%  & 96.8 \% & 22.7 \% & 76.3 \% \\
\hline
\end{tabular}
\end{table}

As the value of $\beta$ separating the spherical and deformed wells we have considered several points: the zeroes of the $2^+_2$ and $0^+_2$ wave functions, and the middle of the barrier.
The results obtained are not very sensitive to this choice. As we see from Table~\ref{table:two} the wave function of the $I^\pi = 0^+_{1,2}$  states are practically concentrated in one well: spherical for the $0^+_1$ state and deformed for the $0^+_2$ state. Qualitatively the same situation is realized in the case of the $I^\pi = 2^+_{1,2}$ states. However, in this case the distribution of the  wave functions between the spherical and deformed wells is less asymmetric.
As it is seen from Table~\ref{table:two} the wave function of the $2^+_1$ state is also located mainly in the spherical well like the $0^+_1$ state. However, its weight in the deformed minimum is equal to 22.7 \%, i.e. significantly larger than for the $0^+_1$ state.

       In addition to the states considered above there are the known 4$^+_1$ state which decays by a strong E2 transition to the 2$^+_2$ state and the 0$^+_3$ state also decaying to the 2$^+_2$ state by a strong E2 transition~\cite{Witt}. This means that both 4$^+_1$ and 0$^+_3$ states can be considered in the quadrupole-collective model presented in this paper. Our preliminary calculations with the potential fitted above have shown that the wave function of the 4$^+_1$ state is concentrated in the deformed well. Therefore, this state can be considered  as a member of the band including $0^+_2$ and 2$^+_2$ states. However, we have obtained that the ratio $[E(4^+_1)-E(0^+_2)]/[E(2^+_2)-E(0^+_2)]$ is equal to ~2.5. This value is much lower than the typical rotational ratio 3.33, however, is larger than the experimental value. The reason is a closeness of the energy of the $4^+_1$ state to the barrier height.
			
			For the energy of the $0^+_3$ state we obtained a value which is much larger than the experimental one. Probably, this means that the structure of the $0^+_3$ state is related to an excitation of the $\gamma$-mode.

\section{Conclusion}

We have studied a possibility to describe the properties of the low-lying collective states of $^{96}$Zr basing on the Bohr collective quadrupole Hamiltonian in terms of axially symmetric shape coexistence.
The potential energy of this Hamiltonian is fixed  to describe the experimental data in the best way. This potential has two minima - spherical and deformed separated by a barrier. Good agreement with the experimental data is obtained for  excitation energies, $B(E2)$, and $B(M1)$ values. It is shown that the experimental value of the excitation energy  of the $2^+_2$ state can be reproduced only if the rotational inertia coefficient in the region of the deformed well is taken five times smaller than in the region of the spherical well.

The calculated value of the $\rho^2(0^+_2\rightarrow 0^+_1)$ is six times smaller than the measured value. This indicates, possible influence of the pairing vibrational mode which is not included in the present consideration.

The calculated value of  the $B(M1; 2^+_2 \rightarrow 2^+_1$) strength demonstrates an excellent agreement with the experimental value. However, this result has been obtained due to modification of the $M1$ transition operator which takes into account the result of the shell model for the g-factor of the spherical configuration.

\section{Acknowledgements}

\begin{acknowledgements}
We thank A. Leviatan, V. Werner, and T. Beck for discussions. This work was supported by the
German DFG under grant No. SFB 1245 and by the BMBF under grant No. 05P18RDFN1.
\end{acknowledgements}

\end{document}